\documentclass[twocolumn,showpacs,aps,prb]{revtex4}

\usepackage{epsfig}
\usepackage{amsmath}

\begin{document}

\title{Scale free property and edge state of Wilson's numerical renormalization group}
\author{Kouichi Okunishi and Tomotoshi Nishino$^{1}$}
\affiliation{Department of Physics, Faculty of Science, Niigata University, Niigata 950
-2181, Japan.\\
$^1$Department of Physics, Graduate School of Science, Kobe University, Kobe 657-8501, Japan}

\date{\today}
\begin{abstract}
We discuss the scale-free property of Wilson's numerical renormalization group(NRG) for the Kondo impurity problem.
The single-particle state of the effective Hamiltonian with a cutoff $\Lambda$ is described by the wavepacket basis having the scale free property;
The energy scale of the system can be controlled by the lattice translation of the wavepacket basis with no reference of rescaling of the lattice space.
We also analyze the role of the Kondo interaction in the context of wavepacket basis and then discuss the scaling and renormalization of the Kondo coupling.
In addition, we clarify the role of the edge state in the lowest energy scale of Wilson NRG.
\end{abstract}

\pacs{05.10.Cc, 75.20.Hr}

\maketitle

\section{introduction}

 Renormalization group(RG) has been one of the most fundamental concept in statistical physics to explain hierarchical structure of the energy scale in nature.\cite{RG}
In particular, such a real-space RG approach as block-spin transformation provides an intuitive view for coarse graining of the fluctuation field and rescaling of the length unit;
It is well-known that the block-spin transformation for the triangular lattice Ising model is often discussed as a textbook example\cite{RSRG}.
For the low-dimensional quantum RG, however, the situation is different;
The conventional block-spin transformation often fails in extracting the low-energy physics of the one-dimensional(1D) quantum system, and improvement of the real-space RG has been an attractive frontier in the quantum physics.
In modern view, this is mainly because the block-spin transformation breaks the quantum entanglement associated with the large quantum fluctuation.\cite{mera,mera2}

 Historically,  two real-space RG approaches not based on the block-spin transformation made great success in the 1D quantum physics:
Wilson's numerical renormalization group(NRG)\cite{wilson,KWW,wrgreview} for the Kondo impurity problem\cite{kondo} and density matrix renormalization group(DMRG)\cite{white,scholl}.
An interesting point is that these two methods have completely different characters.
 Wilson NRG can directly deal with the metallic system with a magnetic impurity, through mapping to the effective 1D chain with a boundary.
On the other hand, we should remark that DMRG includes no scale transformation and no concept of the running coupling constant.
This implies that  DMRG should be considered as a variational optimization method for the matrix product wavefunction\cite{baxter,fannes,ostlund,nishino,scholl2}, rather than a RG based on the coarse graining and the scale transformation.
Thus, Wilson's pioneering work has been the peculiar RG method that succeeds in such energy-scale control of the massless excitation.
 What is its key mechanism?
Answer to this question provides an essential insight for the quantum RG.
Here, we just note that, recently, the entanglement renormalization interestingly reproduces the correct critical index for 1D quantum systems\cite{mera1d}.

In this paper, we will show that Wilson NRG has the scale free property, which is the key of systematical selectivity of the arbitrary energy scale around the Fermi surface. 
For this purpose,  we directly focus on the relation between the  wavefunction and the cutoff parameter $\Lambda$ due to the log-discritization.
In actual NRG iterations, a free fermion site is added step by step with decreasing the energy scale by  $\Lambda$ and  the effective 1D tight-binding model having the exponential modulation of the hopping parameter is eventually generated.
The qualitative role of this cutoff can be viewed as a regulator for the infrared divergence, which introduces effective length scale $\xi\sim 1/\ln\Lambda$.\cite{oku1}
In actual numerical computations,  however, the cutoff value is usually  adopted to be $\Lambda=1.5\sim 2.5$,  which is beyond perturbation level to the original Kondo problem($\Lambda=1)$.
The success of Wilson NRG with the nonperturbative value of $\Lambda$ suggests that the Wilson's cutoff scheme should involve some plausible mechanism to maintain the nature at $\Lambda=1$;
We clarify that the one-particle eigen-wavefunction is modified from the plane wave into the wavepacket by the cutoff, and its lattice translation explains the selectivity of the energy scale.
We also find that there is an edge state in the lowest energy scale near the Fermi surface.
On the basis of the  wavepacket basis above, we perform the exact diagonalization of the Wilson effective Hamiltonian including the Kondo impurity.
We then discuss that the scale factor of the groundstate wavefunction should be renormalized from the value of the non-interacting wavepacket by the Kondo interaction.

This paper is organized as follows.
In section II, we define the effective tight-binding model and then explain its scale free property.
In section III, nature of the wavepacket is analyzed.
In section IV, we resolve the role of the Kondo iteration in the context of the wavepacket basis. 
Section V is devoted to the summary  and discussion.

\section{model and scale free property}

In Wilson NRG, the essential point is to consider the 1D lattice fermion model mapped from the degenerating free electrons around the Fermi surface.
We thus start with the Wilson-type Hamiltonian of the spinless free fermions with the exponentially modulated hopping
\begin{equation}
 {\cal H}_\lambda =\sum_{n=1}^{N-1}  e^{\lambda n }(c^\dagger_{n+1}c_n + c^\dagger_{n}c_{n+1}),
\label{wilsonchain}
\end{equation}
where $c_n$ is a fermion annihilation operator at $n$th site and $N$ denotes the number of sites.
We have also introduced  $\Lambda \equiv\exp(\lambda )>1$ for later convenience.
Thus $n=1$ corresponds to the smallest energy scale and $n=N$ does to the impurity site with the largest energy scale\cite{index}.
Although, in the original work\cite{wilson}, the hopping parameter has a supplemental coefficient and $n$ in $e^{\lambda n }$ term takes a half integer, the essential physics is the same as Eq. (\ref{wilsonchain}).

Let us write the one-particle state as $|\psi\rangle=\sum_n\psi(n)c^\dagger_n |0\rangle$.
Then one-particle Schr\"odinger equation in the bulk region is 
\begin{equation}
  e^{-\lambda }\psi(n-1)+ \psi(n+1)= E e^{-\lambda n} \psi(n) .\label{eigenvalueeq1}
\end{equation}
Note that Eq. (\ref{eigenvalueeq1}) is invariant under the transformation, $\psi(n) \to (-1)^n \psi(n)$ and  $E\to -E$,  which clearly represents the particle-hole symmetry.  
Thus, we basically consider the positive energy solution.

Since the system has no explicit translational symmetry, we employ numerical diagonalization of Eq. (\ref{eigenvalueeq1}) rather than the usual Fourier analysis, for a finite but sufficiently large system.
We assume the free boundary condition and thus what we deal with is the tridiagonal matrix. 
Figure 1 represents the absolute value of the one-particle spectrum for $\lambda=0.1$ and $N=200$, where $j$ indicates the label of the eigenvalue in increasing order.
The Fermi surface is located between $j=100$ and 101, and $j\le 100$ represent negative energy eigenvalues.
Thus, the parity in Fig. 1 corresponds to the particle-hole symmetry.
We also show the amplitude of the wavefunctions corresponding to $j=101$, 130 and 160th eigenvalues, in Fig. 2. 

The most important behavior in Fig. 1 is that, as was already mentioned in Ref. \cite{wilson}, $E$ basically  exhibits the exponential dependence $E\propto \pm \exp(\lambda j)$.
We call this region of the exponential dependence as ``bulk'', since the corresponding wavefunctions are localized in the bulk region of the chain, as can be seen for $j=130$ and 160 in Fig.2.
On the other hand,  we can see that some eigenvalues near the Fermi surface $j\sim 100$ deviate from the bulk lines.
We call these states as ``edge states'', since they correspond to  the edge modes near the Fermi surface, as is the wavefunction of $j=$101 in Fig. 2.

\begin{figure}[t]
\epsfig{file=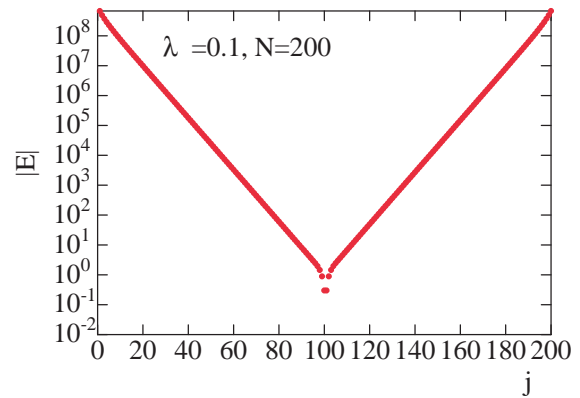,width=7.4cm}
\caption{(Color online)One-particle eigenvalue spectrum of Eq. (\ref{eigenvalueeq1}).
The horizontal axis $j$ indicates the label of the eigenvalue in increasing order.}
\end{figure}
\begin{figure}[th]
\epsfig{file=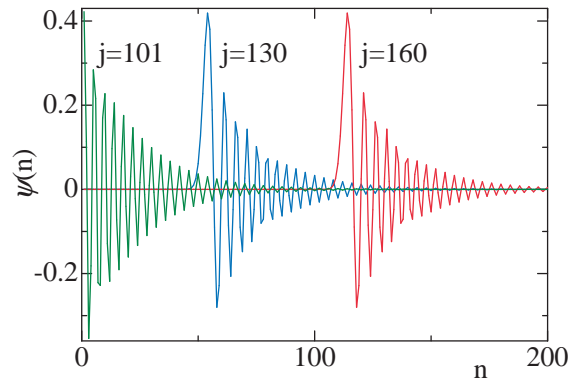,width=7.4cm}
\caption{(Color online)One-particle wavefunction $\psi(n)$ of $j=101$, 130 and 160. 
The horizontal axis means the cite index $n$.
The bulk states $\psi(n)$ of $j=130$ and 160 can be overlapped with each other by a lattice translation, 
while $\psi(n)$ of $j=101$ is the edge state.
}
\end{figure}

We analyze the bulk part of the spectrum in detail.
In connection with the exponential dependence of the eigenvalues, an essential information can be found in the bulk wavefunctions for $j=130$ and 160 in Fig. 2.
The primary notable point is that these wavefunctions have a very similar wavepacket-like shape,  in contrast to the usual plane wave for the uniform chain.
The localization of the wavefunction can be qualitatively understood as follows. 
If a particle carrying a certain energy goes in the larger $n$ region, the particle can not excite the larger energy bonds, into which the wavefunction can not penetrate.
While the particle goes to the smaller $n$ region, the bond of the smaller coupling can not carry the total energy of the particle and thus the wavefunction decays very rapidly.

Another important property of the bulk wavefunctions is that they can be overlapped with each other by the lattice translation;
In Fig. 2, the wavefunctions of $j=$130 and 160  have the very similar shape. 
Indeed,  we can verify that overlap integral of the two wavefunctions after the lattice translation is unity within the computational accuracy.
In order to see this property in analytic level, we introduce 
\begin{equation}
\psi(n)\equiv e^{-\lambda n /2} \phi(n). \label{psidecay}
\end{equation}
and then rewrite Eq. (\ref{eigenvalueeq1}) as 
\begin{eqnarray}
  \phi(n-1)+ \phi(n+1)= \bar{E} e^{-\lambda n} \phi(n) ,\label{eigenvalueeq2}
\end{eqnarray}
where $\bar{E}= E e^{\lambda/2}$.
Now let us consider the effect of the lattice translation.
By shifting $n=n'+m$ and rewriting $\phi(n'+m)\to \phi(n')$,
we obtain
\begin{eqnarray}
&  \phi(n'-1)+ \phi(n'+1)= \bar{E} e^{-\lambda m}e^{-\lambda n'} \phi(n'), \label{eigenvaleeq3}
\end{eqnarray}
which becomes identical to Eq. (\ref{eigenvalueeq2}) by redefining $\bar{E}'=\bar{E}e^{-\lambda m}$.
In other word, the lattice translation of the wavefunction  has one-to-one correspondence to the eigenstate of the $e^{-\lambda m}$-scaled energy eigenvalue, which has been mentioned in Ref.\cite{ueda} in the operator level.
Thus, once a certain energy eigenvalue is obtained, the other eigenvalues and the corresponding eigenvectors can be obtained by the lattice translation combined with $E' = e^{-\lambda m}E$.
In addition, we note that, if $\lambda$ is sufficiently small, we can always find a certain $m$ such that  $E'$ is unity, except for $E$ being exactly zero. 
This implies that the Wilson Hamiltonian is {\it energy-scale-free theory} and thus the problem can be physically well-defined, although the Hamiltonian includes very wide range of the energy scale, $\Lambda \sim \Lambda^N$.

\section{wavepacket basis}

We analyze features of the bulk wavepacket itself. 
For this purpose, we concentrate on the positive energy solution of Eq. (\ref{eigenvalueeq2}) and  consider its ``continuum limit'' $\lambda \to 0$.
Writing $x=\lambda n$, we have
\begin{eqnarray}
&   \lambda^2  \phi''( x) + (2- \bar{E}e^{-x}) \phi(x) =0 , \label{eigenvalueeq7}
\end{eqnarray}
where we have omitted ${\cal O}(\lambda^3)$.
Note that the translation $x\to x+ a$ combined with $ \bar{E}e^{-a} \to \bar{E}$ keeps Eq. (\ref{eigenvalueeq7}) invariant.
Thus we can set $\bar{E}=1$ in the following analysis.
It should be also remarked that the continuum limit $\lambda \ll 1$ can be taken with no reference to the lattice space.

Introducing $y\equiv \exp(-x/2)$, we can convert Eq. (\ref{eigenvalueeq7})  to the modified Bessel type equation,
\begin{equation}
\left[ \frac{d^2}{dy^2} + \frac{1}{y}\frac{d}{dy}  + \frac{4}{\lambda^2} (\frac{2}{y^2} -1 ) \right] \phi(y)=0 .
\label{eigenvaleeq8}
\end{equation}
Assuming that the boundary is at infinity with respect to $x$, and taking account of the normalizability of the wavefunction in $x\ll -1$($y\gg 1$),  we obtain
$\phi(y)=   {\rm const } \times K_{i2c}(cy)$ with $c\equiv \sqrt{2}/\lambda$,\cite{rc} 
where $K_{i\nu}(z)$ is the modified Bessel function of the imaginary order $\nu$.
Using the formula $ \int_0^\infty dz z K_{-i\nu}(z)K_{i\nu}(z) =  \frac{\pi \nu}{2\sinh \pi\nu} $,
we obtain the eigenfunction of Eq. (\ref{eigenvalueeq1}) as
\begin{equation}
\psi(x)=  \sqrt{\frac{\sinh 2 \pi c}{2 \pi c   }}c e^{-x/2} K_{i2c}(ce^{-x/2}).
\label{psisolution}
\end{equation}

The behavior of this solution can be resolved in special limits.
When $x\gg1$($y\ll 1$), we can easily see that Eq. (\ref{eigenvalueeq7}) reduces to the uniform tight binding model, which yields
\begin{equation}
\phi(x) \sim \exp(-i c x).
\label{rtail}
\end{equation}
This is consistent with the asymptotic behavior of the modified Bessel function $K_{i\nu}(z) \sim 2^{i\nu-1}\Gamma(i\nu)z^{-i\nu}  $ in the $z \to 0$ limit.
Thus, the wavefunction $\psi(x)$ exponentially decays  in $x\gg1$ with the oscillation. 
This behavior is consistent with the numerical result in Fig. 2, where we have checked that the tail of the wavepacket decays as $e^{-\lambda n /2}$.

In the limit $z \gg 1$, the asymptotic form is given by $K_{i\nu}(z) \sim  \sqrt{\frac{\pi}{2z}}\exp(-z) $.
For $x\ll 0$($y\gg 1$), thus, we can see
\begin{equation}
\phi(x) \sim e^{x/4}\exp(-ce^{-x/2}) ,
\end{equation}
which decays very rapidly in $x<0$.
This is also in good agreement with the numerical result in Fig. 2.
From these, we can see that the localization of the wavepacket is  basically good and the width of the packet is exactly described by the decay rate of the  high-energy tail, $\xi \equiv 2/\lambda (= 2/\ln \Lambda)$, because of Eq. (\ref{psidecay}).
Also this good localization of the packet justifies that the effect of the boundary can be negligible in the bulk region.

\begin{figure}
\epsfig{file=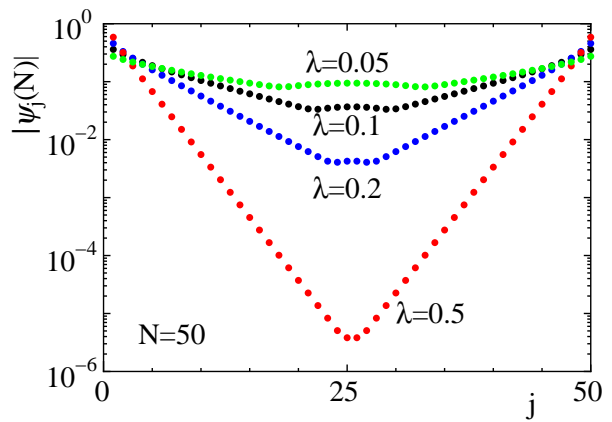, width=8cm}
\caption{(Color online)Amplitude of the  wavefunction at the impurity site $\psi_j(N)$ of $N=50$ for $\lambda=0.05$, 0.1, 0.2 and 0.5, where $j$ indicates the label of an eigenstate. 
We can see the clear exponential dependence on $j$ for the bulk wavefunction.
The small hump around Fermi surface($j\sim 25$) corresponds to the edge states of the lowest energy scale. 
The hump region of the edge states extends, as $\lambda$ becomes small.}
\label{figpsiedge}
\end{figure}

Since the Kondo impurity is located at the right edge of the chain, the amplitude at $n=N$ is of particular importance.
Here we return to the lattice problem and write the index of the wavepacket as $j$. 
Figure \ref{figpsiedge} shows the absolute value of $\psi_j(N)$, for which we can see the exponential dependence of $\psi_j(N)$ with respect to the index $j$.
Because of the resolution of the figure, we plotted the case of $N=50$.
Note that  the particle-hole symmetry of (\ref{wilsonchain}) gives $\alpha_j= -\alpha_{N-j+1}$ with $j\le N/2$($N=$even), which can be also confirmed in Fig. \ref{figpsiedge}.
According to  Eq. (\ref{psidecay}),  the tail of the single fermion wavepacket is described by the damping factor $e^{-\lambda n/2 }$, when  the packet is sufficiently  away from the impurity.
Thus, we can write the amplitude of the wavefunction at the impurity site as 
\begin{equation}
\psi_j(N) =\left\{
\begin{array}{ll}
 \alpha_j \exp(-\lambda j ) & {\rm for}\; 1 \le j \le \frac{N}{2}  ,\\
 \alpha_j \exp( \lambda (j-N-1) ) & {\rm for}\; \frac{N}{2}+1 \le j \le N ,
\end{array}
\right. 
\label{rightedge}
\end{equation}
where $\alpha_j$ is a coefficient of order of unity for $j$-th wavepacket.
Note that $\alpha_j$ is a constant in the bulk region.
However, we should also remark that, as $\lambda$ becomes small,  the small hump region extends around Fermi surface($j\sim 25$); 
The states in the hump region correspond to the edge states of the lowest energy scale, which will be discussed in the next section.

\section{edge state}

\begin{figure}[b]
\epsfig{file=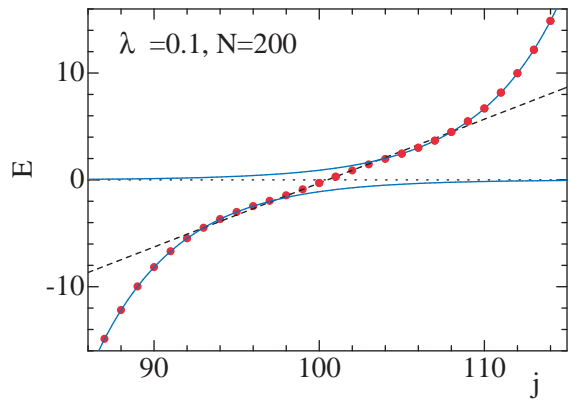,width=7.4cm}
\caption{(Color online)The linear scale plot of the eigenvalue spectrum around the Fermi surface for $\lambda=0.1$ and $N=200$. Solid symbols denote the $j$th eigenvalue. 
The Fermi surface corresponds to $E=0$, which is indicated by the horizontal dotted line.
Solid curves represent exponential dependence of the eigenvalues in the bulk region and the broken line indicates the edge mode (\ref{edgeenergy}).
}
\label{figedgeev}
\end{figure}

In this section, we would like to discuss the boundary effects in Wilson NRG, which can be seen around the Fermi surface($E=0$) corresponding to the left edge of the chain.
Figure \ref{figedgeev}  shows the magnification of the eigenvalue spectrum around $E=0(j\sim 100$), where the scale of the vertical axis is linear.
As was mentioned, the energy eigenvalue distribution in the bulk part has the exponential dependence, which is represented as broken curves.
However, we can see that the spectrum in the vicinity the Fermi surface deviates from the exponential dependence and becomes rather linear with respect to index $j$. 
These states have the lowest energy scale near the Fermi surface, which implies that the property of the edge state is important.
Indeed we can see that the wavefunction of $j=101$ in Fig. 2 localizes at the edge region.
Although the shape of this wavefunction itself is similar to the right side of the bulk wavepacket, the eigenvalue distribution exhibits different structure from the bulk region.

In order to analyze the edge states, we consider another continuum limit, which may be similar to the usual approach for the fermion with the linear dispersion.
We rewrite $\phi(n)= e^{\pm i \pi n/2}\eta(n)$, where the phase factor basically corresponds to the right or left going Fermi wave numbers of the uniform case.
Then, equation (\ref{eigenvalueeq2}) becomes
\begin{equation}
\mp i[\eta(n-1)-\eta(n+1)] = \bar{E} e^{-\lambda n} \eta(n).
\end{equation}
Writing $x=\lambda n$ again and maintaining the leading term of ${\cal O}(\lambda)$, we have 
$(2i \lambda \frac{d}{dx} \mp \bar{E} e^{-x})\eta(x) =0.$
The solution of this equation can be easily obtained as
\begin{equation}
\eta(x)=C \exp(\mp i\frac{\bar{E}}{2\lambda}e^{-x}) ,
\label{etasolution}
\end{equation}
where $C$ is a certain constant.
For $x \gg 1$, $\eta \to $const and thus $\psi \sim e^{-x/2}e^{ \pm i \pi x/2\lambda}$. 
For $x \ll -1$, however, $\eta$ becomes highly oscillating, and if its period becomes smaller than the lattice space, this continuum limit breaks down.
Moreover, the wavefunction $|\psi(x)|^2 \sim e^{-x}  \to \infty$ in $x \to -\infty$.
In this sense, the wavefunction (\ref{etasolution}) is prohibited in the bulk limit.
If the system has the left edge, however, such a breakdown can be avoided. 
When the boundary condition is $\psi(0)=\psi(N+1)=0$, the energy eigenvalue is determined as 
\begin{equation}
E=2\pi \lambda e^{-\lambda/2}q,
\label{edgeenergy}
\end{equation}
where  $ q=\pm 1/2, \pm 3/2 \cdots$\cite{integer} is a quantum number measured from the Fermi surface.  
In fact, we have $E=0.2988$ for $\lambda=0.1$ and $q=1/2$, which is in good agreement with the numerical result $0.2989$ in Fig.\ref{figedgeev}. 
As $\bar{E}$ increases,  the oscillation of Eq. (\ref{etasolution}) becomes significant and then the wavefunction crosses over to the bulk wavefunction (\ref{psisolution}), away from the edge.  
Since the phase difference between $n=0$ and $1$ in Eq. (\ref{etasolution}) is about $\bar{E}/2$, the criterion of the crossover is estimated as $E \simeq \bar{E}\sim 2 $, which is also consistent with Fig. \ref{figedgeev}.

Although the lowest energy behavior is described by the boundary excitation (\ref{edgeenergy}), the role of the edge state attracts less interest so far.
One reason for this is that the practical value of the cutoff is $\Lambda^2 \simeq 1.5 \sim 2.5$, for which the level space becomes too sparse to detect the linear region in the spectrum. 
Here, it should be noted that the generalized Wilson NRG scheme with much smaller $\Lambda(=1.02)$ exactly extracts the low-energy spectrum of the critical XXZ chain.\cite{oku1}
Another reason is that, as in Fig. \ref{figedgeev},  the edge wavefunction smoothly varies its shape into the bulk wavepacket and the tail of the edge wavefunction shows the same dumping  $|\psi|^2\sim e^{-x}$ as the bulk wavefunction.
Thus, one can increase number of sites seamlessly  in NRG iterations, with no explicit reference of the edge state.
Finally, we make a comment on the right edge corresponding to the highest energy scale;  In contrast to the left edge above,  the universal eigenvalue structure can not be seen for the right-edge states.

\section{effect of impurity spin}

On the basis of the nature of the wavepacket, let us discuss the effect of the Kondo impurity in Wilson NRG. 
Assume an impurity spin $\vec{S}$ with the Kondo coupling $J_K$ at the right edge of the chain($n=N$).
The free electron Hamiltonian ${\cal H}_\lambda$ is diagonalized in the wavepacket basis, ${\cal H}_\lambda=\sum_{j\sigma} E_j f^\dagger_{j\sigma}f_{j\sigma}$, where $f_{j\sigma}$ is a fermion operator corresponding to the $j$-th eigenvalue of (\ref{wilsonchain}), and the spin index $\sigma$ is noted explicitly.
In the following, the eigenstate described by $f_{j,\sigma}$ is called as ``orbital electron'' for convenience.
As was discussed in Sec. II, $E_j \propto \exp(2 \lambda j)$ for $1\le j \le N/2$, and $E_j\propto \exp(\lambda(N-2 j))$ for $N/2+ 1\le j \le N$\cite{factor2}.
On the other hand,  the matrix element associated with the impurity  is determined by the tail of the wavepacket at the right edge,
\begin{equation}
{\cal H}_{\rm imp}=  J_K e^{\lambda N} \sum_{i,j,\sigma,\sigma'} \psi^*_i(N)\psi_j(N)  f^\dagger_{j\sigma}[\vec{\sigma} \cdot \vec{S} ]_{\sigma,\sigma'} f_{j\sigma'}, \label{kondop}
\end{equation}
where $\vec{\sigma}$ is the $S=1/2$ spin(not Pauli) matrix associated with the orbital electron, $\vec{S}$ represents the impurity spin of $S=1/2$, and $\sigma$ takes $\sigma \in \uparrow,\downarrow$ .
Here, it should be recalled that, in Wilson NRG,  the energy scale of the Kondo coupling itself is scaled up as $J_K e^{\lambda N}$, in the process of the iterative increase of the free electron sites.

 It is instructive to write the matrix elements of ${\cal H}_{\rm imp}$ in the orbital bases.
Using Eq. (\ref{rightedge}), we have the diagonal parts with respect to the spin as,
\begin{eqnarray}
\langle  i,\sigma,S |{\cal H}_{\rm imp}| j,\sigma,S\rangle = \pm\frac{J_K}{4} \alpha_i \alpha_j  C_{i,j} ,\label{diagonal}
\end{eqnarray}
where 
\begin{eqnarray}
C_{i,j}\equiv
 \left\{
\begin{array}{ll}
\displaystyle  e^{\lambda(N-i-j)}, &  1 \le i,j \le \frac{N}{2} \\
\displaystyle  e^{\lambda(-i+j-1)}, & 1 \le i\le \frac{N}{2},\;  \frac{N}{2}+1 \le j \le N\\
\displaystyle  e^{\lambda(i-j-1)}, & \frac{N}{2}+1 \le i \le N,\; 1 \le j \le \frac{N}{2} \\
\displaystyle  e^{\lambda(-N+i+j-2)}, &  \frac{N}{2}+1 \le i,j \le N 
\end{array}
\right. .
\end{eqnarray}
Also, the off-diagonal part is given by
\begin{eqnarray}
\langle  i,\uparrow,\Downarrow |{\cal H}_{\rm imp}| j,\downarrow,\Uparrow\rangle &=&\langle  i,\downarrow,\Uparrow |{\cal H}_{\rm imp}| j,\uparrow,\Downarrow\rangle \nonumber\\
&=& \frac{J_K}{2} \alpha_i \alpha_j C_{i,j} ,\label{offdiagonal}
\end{eqnarray}
where $\Uparrow,\Downarrow$ represent the impurity spin.
Here we should remark that the damping factor of the wavepackets near the Fermi-surface($i,j\simeq N/2$) is canceled out with the scale factor of the Kondo coupling.
This implies that the fermionic $2k_F$-oscillation of Eq. (\ref{rtail})  is correctly recovered.
Thus, the phase shift of the Fermi-level state due to the impurity scattering can be correctly taken into account, although the wavepackets are resolved with respect to the energy scale.

The eigenvalues of the impurity term ${\cal H}_{imp}$ are just given by $-3e^{\lambda N}J_K/4 $ for the singlet,  $e^{\lambda N}J_K/4$ for triplet and the zero energy with the huge degeneracy.
By the perturbation of ${\cal H}_\lambda$, the most of the degenerating zero eigenvalus are lifted to have a ``dispersion'',  whose maximum energy scale is of order of $E_N\sim e^{\lambda N}$.
For $J_K > 1$,  we can naively treat ${\cal H}_{\lambda}$ as a perturbation against to  ${\cal H}_{imp}$, since ``band width'' of the free orbital electrons is smaller than the singlet-triplet gap of  ${\cal H}_{imp}$.
In the context of physics, an orbital electron localizes at the edge and forms a singlet or triplet pairs with the impurity spin. 
This is the local Fermi liquid fixed point.

For the weak coupling case($J_K \ll 1$), the dominant part of the Hamiltonian is described by ${\cal H}_\lambda$, and  ${\cal H}_{\rm imp}$ is a perturbation for small $N$.
As is well known, however, the low-energy physics should flow to the strong coupling limit for sufficiently large $N$.
For simplicity,  we assume $1 \le i,j \le \frac{N}{2}$. 
For $2M$-site lattice translation\cite{factor2}, then, the scale factor $C_{i,j}$ in (\ref{diagonal}) and (\ref{offdiagonal})  is scaled by $ C_{i+M,j+M} \exp(2\lambda M )$, while $\alpha_{i,j}$ in the bulk region is independent of $i,j$.
Thus we can read  $J_K e^{\lambda N} \psi^*_i(N)\psi_j(N) = \tilde{J}_K e^{\lambda N} \psi^*_{i+M}(N)\psi_{j+M}(N)$ with $\tilde{J}_K\equiv J_K \exp(2\lambda M)$.
The role of Kondo coupling of a $J_K$ at a certain energy scale $e^{\lambda j}$ can be viewed as that of the scaled Kondo coupling $J_K \exp(2 \lambda M )$ in a higher energy scale $e^{\lambda(j+M)}$.
In the following,  we call this factor $\exp(2 \lambda M )$  as ``canonical scale factor'' of the Kondo coupling.

\begin{figure}[h]
 \epsfig{file=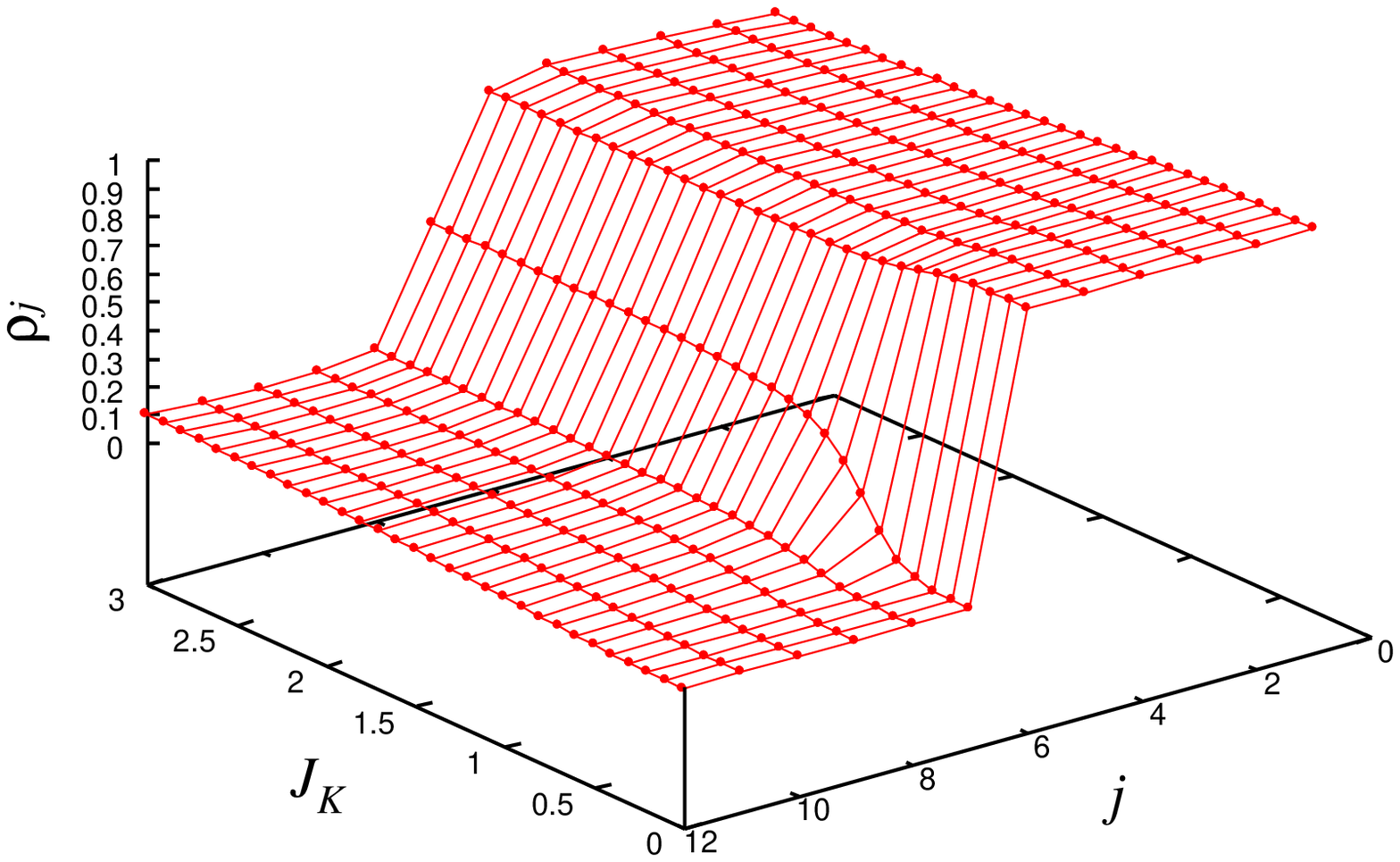,width=7.2cm}\\
~\\
\epsfig{file=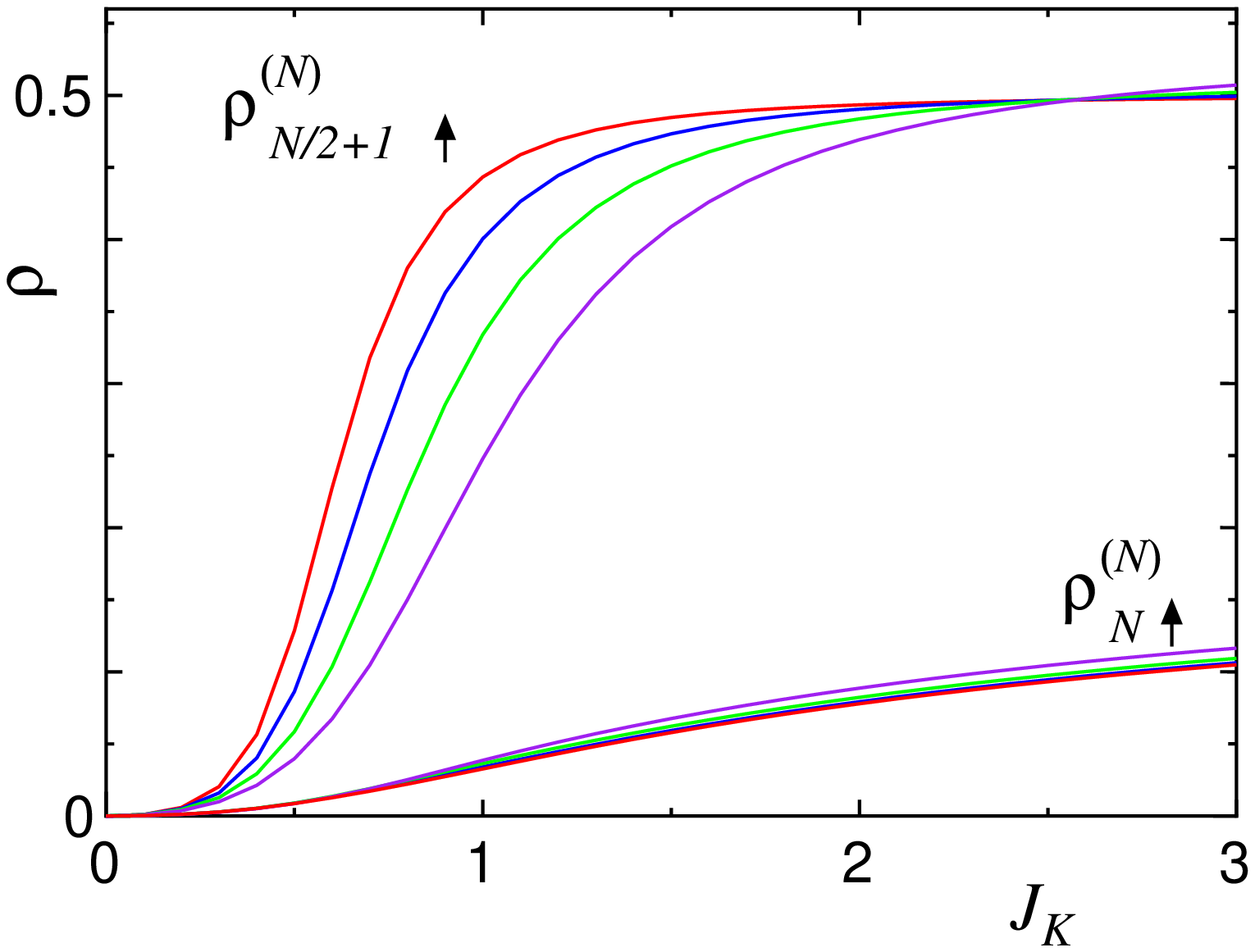,width=7cm}
\caption{(Color online) (a) Kondo coupling dependences of  $\uparrow$-electron density for $N=12$ with $\lambda=\ln 2 /2 \simeq 0.347(\Lambda=\sqrt{2})$, where $j$ denotes the index of the orbitals.  
The Fermi surface of the free orbital electron corresponds to $j=6.5$.
(b) Size dependences of $\uparrow$-electron density for $\lambda\simeq 0.347(\Lambda=\sqrt{2})$.
The curves  indicate the $\uparrow$-electron density at the Fermi surface  $\rho^{(N)}_{N/2+1 \uparrow}$ for $N=6$, 8, 10, and  12 from down to up. 
The solid curves of $\rho^{(N)}_{N\uparrow}$ represent the $\uparrow$-electron density at the highest energy scale orbital.
}
\label{figorbital}
\end{figure}

\begin{figure}[h]
  \epsfig{file=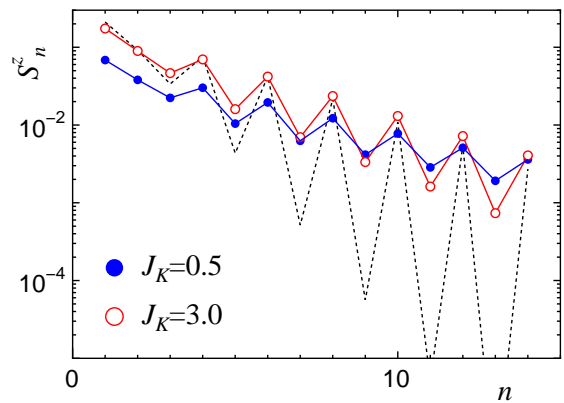,width=7.5cm}
\caption{ 
(Color online)Spin distribution  in the real-space representation for $N=14$ and $\lambda=\ln2/2$.
The open(solid) circle indicates $J_K=3.0$(0.5). 
$n$ denotes the lattice index, where $n=14$ corresponds to the impurity site.
The dotted line means the spin distribution of the free orbital of $j=8$.
}
\label{figdn}
\end{figure}

The above scaling argument might suggest that the low-energy physics basically flows to the strong coupling limit by the canonical scale factor $e^{2\lambda M}$.
However, the actual spectral flow is rather complicated;
Since ${\cal H}_\lambda$  has the same scale factor as ${\cal H}_{imp}$ as well, the competition of ${\cal H}_\lambda$ and ${\cal H}_{imp}$ yields the nontrivial Kondo energy scale.
Of course, it is well established that the entanglement between the impurity and orbital electrons rapidly develops around the Kondo energy scale given by $\exp(-1/2J_K)/\sqrt{2J_K}$.
In the present context, this suggests that the site dependence of the effective Kondo coupling should be deviated from the naive canonical scale factor.

In order to see the effect of the Kondo interaction, we perform exact diagonalization of the Wilson's effective Hamiltonian up to 14 orbitals with the subspace of electron number $N$ and total-$S^z=1/2$(including the impurity spin). 
We use $\lambda=\ln 2 /2 \simeq 0.347(\Lambda=\sqrt{2})$ in order to see clearly the role of the impurity within the exact diagonalization.
Then, we focus on the nature of the groundstate wavefunction, rather than the spectral flow.
Let us write the $\uparrow$-electron density of the groundstate for $N$-sites system as  $\rho^{(N)}_{j\uparrow}\equiv \langle f^\dagger_{j\uparrow}f_{j\uparrow}\rangle$, where $j$ indicates the orbital index.
Figure \ref{figorbital}(a) shows the $J_K$-dependence of  $\rho^{(N)}_{j\uparrow}$.
Note that, for $N=12$,  the Fermi surface of the free electron is located between $j=6$ and 7.
As $J_K$ increases, the $\uparrow$-electron density of $j=7$ approaches $0.5$, which implies that the impurity spin is screened and then the remaining spin-1/2 chiefly moves to the orbital near the Fermi surface.
At the same time, we can see that  the density for $j\ne 7$ orbitals is also modified from the case of free electrons, even though they are embedded deeply away from the Fermi surface.
This implies that the high energy orbitals are certainly entangled with the impurity spin.

In Fig. \ref{figorbital}(b),  we  show the size dependence of the  $\uparrow$-electron density at $j=N/2+1$ for $N=6$, 8, 10 and 12. 
We can easily see that, as $J_K$ increases,  $\rho^{(N)}_{N+1/2,\uparrow}$  approaches $\rho\simeq 0.5$; 
 the $S^z=1/2$ at the impurity is passed on the orbitals at the lowest energy scale around the Fermi surface, and the number of the orbital sites is effectively reduced to be $N-1$.
Moreover, as $N$ increases with $J_K$ fixed,  $\rho^{(N)}_{N+1/2,\uparrow}$ approaches the value of the strong coupling limit. 
This behavior is of course equivalent to the iterative process of the Wilson NRG.
On the other hand, the $\uparrow$-electron density in the highest energy scale  $\rho^{(N)}_{N\uparrow}$ shows little size dependence, which is almost negligible within the scale of Fig.\ref{figorbital}(b).

In Fig. \ref{figdn}, we show the real-space distribution of the spin density of the groundstate for $N=14$.
Note that the charge distribution of all $J_K$ is uniform in the real space, indicating the spin-charge separation.
The singlet pair of the impurity and the orbital electron itself is invisible to a spin expectation value.
However,  one can extract an essential information from the distribution of magnetization passed on the orbital electrons from the impurity site.
We first discuss the strong coupling case $J_K=3.0$. 
We have confirmed that the spin distribution for $J_K=3.0$ is basically identical to that of $J_K=100$.
The dominant single-particle component in the groundstate is the orbitals around the Fermi surface($j=7$ and 8 for $N=14$).
On the other hand,  the groundstate wavefunction also includes certain amplitudes of the higher energy orbitals, as can be seen in Figs. \ref{figorbital}. 
The  decay rate of the spin density in the strong coupling limit is expected to be close to the canonical value $\lambda =  \ln 2 /2 \simeq 3.47 $,  but to have a certain deviation from it. 
In Fig. \ref{figdn},  the open circle at $n=$even is very similar to the dotted line of the free orbital of $j=8$,  but a precise analysis for the decay rate of the spin density for $J_K=3.0$ at $n=$even sites actually gives $\lambda^*=0.312\simeq 0.9\lambda$.
In addition, we can see that the spin density at $n$=odd sites clearly deviates from the single particle result.
The above results of the $S^z$ distribution  illustrate that the wavefunction in the strong-coupling limit includes the nontrivial correlation effect, where  $\lambda$ is slightly renormalized to $\lambda^*\simeq 0.9\lambda$.
This suggests that the scale factor could be also renormalized to $e^{2 \lambda^* M}$ from the canonical value in  the wavefunction level.

We next turn to the case of $J_K=0.5$, which is in the crossover regime of the Kondo coupling within $N=14$.  
For $J_K=0.5$, the total-$S^z$ of the orbital electron is $0.233$, which implies that the screening rate of the impurity spin is $47$\%.
Then, we can see that the decay rate reduces to be $\lambda^* = 0.215 \simeq 0.62 \lambda$, which illustrates that the  wavefunction in the crossover regime is entangled with the states in the wide range of the energy scale.
In the context of the wavepacket basis, this implies that hybridization of the higher-energy orbitals generates an eigenvalue of the lower-energy scale.
As the system size $N$ increases, we have confirmed that the renormalized $\lambda^*$ approaches the strong coupling limit value $0.312$.
In this sense, the size dependence of $\lambda^*$ might be related with the scaling function of the Wilson NRG(Eqs. VIII.45, 46 or IX.58 in Ref.\cite{wilson}).
For the detailed analysis of $\lambda^*$, the wavefunction of much larger system size is needed beyond the exact diagonalization.

\section{summary and discussion}

We have clarified that the key mechanism of Wilson NRG is the scale free property of the Wilson Hamiltonian (\ref{wilsonchain});
The cutoff $\Lambda$ results in  the wavepacket basis described by Eq. (\ref{psisolution}) and its lattice translation enables for us to control the energy scale of the system with no reference to rescaling of the length unit.
In addition, we have shown that the Hamiltonian (\ref{wilsonchain}) has the edge states at the lowest-energy scale, which has not been mentioned so far.
On the basis of wavepacket, we have performed the exact diagonalization including the Kondo coupling.
We then revealed that the nontrivial effect of the Kondo coupling can be detectable as a renormalization of the decay rate of the groundstate spin density;
The value of $\lambda^*$ is close to the canonical value $\lambda$ in the strong coupling limit, while the renormalization effect becomes large in the crossover regime. 
Since the effective length scale of the ground state wavefunction is given by $\xi^* = 1/\lambda^*$,  it may be efficient to increase the number of the basis in the crossover regime.

In the context of the uniform 1D critical system, the critical phenomena are usually characterized by the power-low decay of physical quantities.
However, the exponential modulation introduces the cutoff of the  infrared divergence and then the criticality of the uniform system would be converted to the nontrivial renormalization of $\lambda^*$.
In particular, the connection of the renormalized scale factor in Wilson NRG  to  the DMRG analysis of the Kondo screening cloud for the uniform chain with a boundary\cite{sorensen} is an interesting problem. 
We also remark that, as in Ref.\cite{oku1},  the regulator for the infrared divergence due to such scale-free modulation of the interaction is a generally applicable idea to the 1D quantum many body system.
For further research, it may be essential to clarify the direct relation between the entanglement and the renormalized scale factor.
Recently, the connection of Wilosn NRG to the matrix product state is pointed out in Refs.\cite{hofstetter,VMPA}.
Moreover, very recently,  the boundary critical phenomena can be correctly captured by the entanglement renormalization.\cite{bmera, bmera2}
We believe that the present result stimulates a new frontier in the quantum RG.

\acknowledgments

This work is supported by Grants-in-Aid for Scientific Research (No.20340096 and No. 22540388). It is also supported by Priority Area ``High Field Spin Science in 100T''.
A part of numerical calculations were carried out on SX8 at  YITP in Kyoto University.

%%%%%%%%%%%%%%%%%%%%%%%%%%%%%%%%%%


\begin{thebibliography}{99}
\bibitem{index} The order of the lattice index is opposite to Ref.\cite{wilson}.
\bibitem{rc} The value of $c$ may be renormalized by higher order terms. 
In the following discussion, $c$ should be a phenomenological parameter.
\bibitem{factor2}  Shift of index $j$ corresponds to the lattice translation of two sites, because of the particle-hole symmetry.  Thus, the energy scale of $E_j$ is scaled with  $\exp(\pm 2\lambda j)$.
\end{thebibliography}
\end{document}